# Classification of Emergency Scenarios

Mathieu Münch
Multimedia Communications Lab, Technische Universität Darmstadt
Rundeturmstr. 10, 64283 Darmstadt, Germany
Email: *mathieu.muench@stud.tu-darmstadt.de*

*Abstract*—In most of today's emergency scenarios information plays a crucial role. Therefore, information has to be constantly collected and shared among all rescue team members and this requires new innovative technologies. In this paper a classification of emergency scenarios is presented, describing their special characteristics and common strategies employed by rescue units to handle them. Based on interviews with professional fire-fighters, requirements for new systems are listed. The goal of this article is to support developers designing new systems by providing them a deeper look into the work of first responders.

## I. INTRODUCTION

Information plays a crucial role in almost every emergency scenario in today's world. It can be the knowledge of a gas container in the basement of a burning house that makes the difference between a successful mission and a disaster.
The collection of information is not a singular process. During an entire rescue mission all members of the rescue team are continuously confronted with changes in situation and facts. All information has to be evaluated, filtered and successfully transmitted to decision makers, deciding on the course of the rescue mission.
The first phase of any emergency scenario is especially chaotic. There is almost no information available. All knowledge has to be collected by exploring the site.
Fast decisions have to be made and predominant chaos has to be overlooked. Therefore, the collection of information as well as its right delivery is the main challenge.

Nowadays collection and transmission of knowledge and information is manly done face to face and by radio communication. Unfortunately the transmission of large data is not possible. Some western countries like Germany, still use an analog transmission. This results in inefficient communication when there is high channel utilization because only one participant at a time can use the channel.
The improvements in mobile networking, sensor and communication technology provide new possibilities to enhance the state of the art communication. [1]
In large scale emergency scenarios multiple organizations with different goals and approaches operate together.
The information has not only to flow among members of the same organizations but also among members of different organizations so that they can cooperate with each other.

To design such systems, a basic knowledge of emergency scenarios and their handling is needed.

This paper endeavors to provide a classification of emergency scenarios and outline their characteristics and risks. Also the effect on victims as well as the involved rescue personnel and their first response methods to the emergency events are described.

The vast diversity in emergency scenarios requires an equally high number of techniques to manage them. The goal of this work is to find similarities between the different scenarios and outline their communication structure.
This paper is based on extensive interviews with two leading fire-fighters of Frankfurt am Main[2] and Darmstadt[3], a leading fire-fighter of the chemical park of Merck[4] and a member of the Red Cross in Germany[5].
They provided a deep look into their best practices and pointed out some important requirements new systems can be incorporated with.
In addition to scientific research paper the German fire-fighter law called "Feuerwehr Dienstvorschrift" is used as a reference.
Section II shows related work that also deals with classification of emergency scenarios and/or requirements for new systems in this domain. In section III, the German emergency organizations and their communication equipment is presented. Section IV deals with the general requirements of new communication systems for emergency scenarios and section V provides a classification of them. Due to the specialization of all four interview partners in everyday emergency scenarios, the mid-scale and large-scale emergency scenarios are covered, less the disaster scenarios. After classifying and describing the different scenarios general similarities are found in section VI and open topics are discussed in section VII. Finally a short conclusion is drawn in section VIII.

## II. RELATED WORK

The classification of emergency scenarios has partially been done by other researchers worldwide. However, the classification has always been specialized to a certain scale or a fewer real-life scenarios. Most classification describe disaster scenarios.
In N. Sanderson's work: Developing mobile middleware - an analysis of rescue and emergency operations"[6] 3 disaster scenarios are described in detail. An earthquake scenario, a train accident located at the Bergensbanen railway in Norway, and a subway station accident in the Paris Metro system.
A detailed description of disasters can also be found in the annual world disaster reports[1] (WDR) of the International Federation of Red Cross and Red Crescent Societies. Especially the WDR 2005 focuses on information in disasters [7].
Classifications and requirements can also be found in firefighter reports and manuals such as the German Feuerwher Dienstvorschrift [8][9] or the Canadian Urban Search and Rescue Classification Guide[2].

## III. GERMAN EMERGENCY ORGANIZATION

The biggest emergency organizations in Germany is the fire Brigade. [10] It consists of 1,039,737 volunteers, 27,816 professionals and 32,752 plant fire-fighters.
The fire-fighters put off fires and carry out rescue and relief operations in road accidents, natural disasters like Storm damage, floods. In bigger cities they also provide the rescue service.
Other major organizations include the Humanitarian NGO's such as the German Red Cross society [11], the Malteser Hilfsdienst [12] and the Johanniter Accident Assistance [13]. They provide rescue services and medical support during large scale scenarios. With its rescue dog unit, the German red cross is also involved in trace and localization missions.
As part of the international Red Cross Society, a large spectrum of the German red cross is also the international relief.
As last part of Germany's emergency organizations the Agency for technical Relief "Technisches Hilfswerk" (THW) has to be mentioned. The THW is a governmental organization, founded to ensure the civil defense in cases of disaster. It now acts also on an international level under the guidance of the German government.
The communication between the rescue organizations is done over radio. Therefore, the frequency around 80 Mhz and 170Mhz are reserved. The communication takes place in the analog domain. Therefore, a channel is always blocked while one participant is transmitting. This becomes a serious problem while an emergency scenario enlarges.
Important messages such as mayday calls could get lost due to interference from other participants.
Moreover, low frequencies do not support high data transmission rates. With a digital usage of the channels, a maximum capacity of 28kbit/s for data in parallel to voice communication can be achieved[14].

The use of other digital communication techniques is not common. For bigger scenarios, the operation control is equipped with a computer connected to the Internet via a mobile connection.

## IV. GENERAL REQUIREMENTS

By developing new support systems for first responders it should always be taken into account that the users are operating under stressful conditions[2]. The system should support them without requiring much user interaction or distracting them with unnecessary data.

The mobility of rescuers requires a *wireless transmission*. Only in a larger scale emergency scenario like a disaster, a cable connection between operation control points may be advantageous [3] because of the high congestion on wireless channels.

A highly dynamic emergency scenario can lead to *fragmentation*. The system should be able to handle the leaving and joining of communication partners as well as splitting and assembling network partitions[6]. During a rescue mission, the affected area is divided into operation zones that are covered by different rescue units. The communication system should be able to represent these zones[3].
The first minutes of an emergency are not only the most important but also the most work-expended, and a basic communication has to be established autonomous without requiring the rescuers to build up an infrastructure. The system should automatically be set up within a short boot time. As the systems are only support systems, their users should not depend on them. functionality[2].
"A system that crashes and has to be reestablished by working fire-fighters will never be accepted."[3]
To be effective, a strict hierarchy among rescue personnel exists. The system should take this hierarchy into account[3].

Since all participants can provide sensitive data, the privacy of rescuer and especially of victims has to be ensured[5].

If a communication system enables certain critical capabilities, a control mechanism has to avoid abuses of this technology. All technical devices have to be shock resistant, splash-proof and if used in hazardous area, heat resistant as well as explosion-proof [4].

## V. CLASSIFICATION OF EMERGENCY SCENARIOS

The common sense of all interviews was that a detailed classification of emergency scenarios is not possible. The unpredictability and differences are too high. Emergencies are results of one or more hazardous events. These events indeed can roughly be classified into:

- Fire, Explosion
- Accident (Car, Train, Machine )
- Chemical, Biological, Nuclear
- Collapse (Building), Burial, Drowning
- Natural disaster (Flood, Storm, Earthquake )

Depending on the size and the impact on life and health of the population, an emergency scenario can be classified as a disaster.

Disaster countermeasures are in the beginning the same as for normal emergency scenarios. The local head of operation can declare the scenario as a disaster and more forces are mobilized. Then, the command structure changes. Depending on the disaster size one operation command stand with its own communication unit is established. It controls the

---
[1] http://www.ifrc.org/publicat/
[2] http://www.publicsafety.gc.ca/prg/em/usar/usar-guide-eng.aspx

multiple local command stands.[15]

The development of new communication systems is primary not dependent on the event itself but on the reaction of the first responders. Therefore, examples of emergency scenarios are presented as well as the typical responses.

It is important to mention, that the following examples do not represent a universal solution but a guideline from which first responders deviate in dependence of events and conditions of the individual emergency scenarios.

The examples are structured in the same way: First the characteristics and risks of an event are described, the involved rescue personnel are listed as well as the potential victims and their characteristics. Finally, the common strategies to tackle the hazardous situation are presented.
An overview of the emergency scenarios can be found in Table III.

*A. Conflagration / Fire*

*1) Characteristics:* A conflagration is an uncontrolled destructive fire that destroys property and threatens life and health of humans and animals. Conflagration ignites predominantly in buildings. Urban areas have a particular high risk of conflagration. A fire in a forest is called forest fire and is fought differently from conflagration in buildings.

A conflagration can lead to new conflagrations in surrounding areas as well as the collapse of burning buildings. By melting cables, a fire can also easily produce an electrical hazard. If not removed or cooled, gas tanks as well as gas pipes produce the risk of explosion[16].

Chemical substances stored or manufactured in burning buildings also bear a substantial risk for further aggravation.

The main danger for humans however is the poisonous smoke. It causes difficulties in breathing, dispossesses oxygen and replaces it with life threatening gases such as Carbon Monoxide that brings unconsciousness as well as death. The non-transparent smoke also lowers the orientation capability. Moreover, dangerous gases do not have to be opaque.

Due to the enormous heat inside a room, a lack of oxygen does not prevent fire from spreading. Inflammable material, such as furniture, dissociates inflammable gases that immediately inflame in an explosion with contact of oxygen. This can happen through bursting window glass because of the heat but also trough fire-fighters opening a door. This "flash over" can in seconds set a room on fire and bring rescuers to life threatening situations.

*2) Rescue personnel:* The main rescuers are fire fighters. A middle size fire of an office building is covered by approximately 3 to 5 groups of 9 fire-fighters. In full action 6 units of 2 fire-fighters are operating inside the building with SCBA's.[17]

In well-equipped areas such as cities, two paramedics per victims are available as long as the number of victims stays low. For high numbers of victims, see: V-C. One to two emergency physicians should also be available.

If the building tends to collapse personnel from the technical relief agency can support the fire-fighters.

*3) Victims:* Victims in conflagration scenarios usually leave the hazardous area[2]. This is supported by evacuation plans and trainings. In private homes or at work, the locality is well known so that disorientation of escaping persons plays a reduced role[4]. It is however the opposite in public buildings such as concert halls, shopping malls or administrative offices. An unclear marking of emergency exits and evacuation routes can lead peoples to be trapped inside the building.

Especially during the night at sleep, smoke cannot be smelled. Victims may wake up and try to escape but often it is too late for them because the poisonous smoke has filled the room. Panic reactions are not rational. Even when most people stay calm some, especially children, tend to hide themselves under objects or in closets. Once the main evacuation of the building is done it is not likely that more people will leave the building. They have to be searched.

*4) Strategies:* After arriving at the incident, the group leader explores the situation. Because he is the first decision maker and operates under a high stress level his decisions will be fundamental for the on-going mission. He has to gather the basic information shown in Chapter X . If humans are in danger the main priority will be rescuing them. The limit for a successful reanimation after a smoke poisoning lies at 17 minutes[18]. The head operator will bring all available forces inside the building. This first phase is chaotic and hindered by insufficient knowledge of the incident. The first arriving leader also has to decide where the vehicles have to be parked, so that they are near the site but do not block any important access roads for later incoming forces.

The rescue units searching the building move slowly. Most of the time they do not have floor plans for navigation and often the smoke is dense so that they have no view. The search process is done by moving through the room using the hands to search for the victims. If available, an infrared camera is used.

The position of the individual rescue units is described verbally over radio transmission to the group leader. He has to keep track of all the movements of his team. After the arrival of a second and third group, a leading structure is established. The incident area is separated into zones. A leading channel is established as well as a communication channel for SCBA units. If required experts are called from the outside and a command stand is established.

*B. Traffic Accident*

*1) Characteristics:* The highest risk traffic accidents pose is the danger produced by ongoing traffic, especially on highways. An ignition and a potential resulting explosion have to be taken into account although they are not the most likely ones to occur. A higher danger comes from loaded goods of damaged transporters such as chemicals and fuel. The load is not always denoted properly, especially if a transporter has loaded lots of individual small packages containing goods that alone do not produce a risk, but they still can in sum have a dangerous impact on the emergency scene. An example could be a transporter full of cigarette lighters.

A special case of traffic accidents is train accidents. A train accident often implies a high number of injured or dead people that are very difficult to reach and rescue. If the train is driven by electricity, the high voltage on the overhead line and inside the trains capacitors can make accessing the inside impossible.

If the train is driven by diesel, high amounts of fuel can leak out. If the accident happens in an urban area, depending on the load, an evacuation of the surrounding area may be inevitable.

The rescuers can also be endangered and hindered by an accident occurred on unstable ground, in mountainous area, in tunnels or on bridges.

*2) Rescue personnel:* In general two groups of nine people provide the material and manpower to cover a small car accident. For larger scale accidents these numbers have to be multiplied.[17]

In well-equipped areas such as cities 2 paramedics per victims are available as long as the number of victims stays low. For higher numbers of victims, see V-C. One to two emergency physicians should also be available.

A train accident with masses of injured people is counted as disaster. The THW supports the fire-fighters with heavy equipment.

*3) Victims:* Victims of accidents are usually injured and under shock. They are often trapped in their vehicles. Nevertheless they do not suffer from smoke poisoning so that the rescue team has more time to work in a more controlled and gentle manner if trapped persons can afford a longer rescue time. Loss of blood is usually the main cause of concern.
Moreover, persons that could escape out of their vehicles can act in unpredictable ways due to shock.

*4) Strategies:* The first action is blocking dangerous traffic around the emergency scene to protect the first responders. Cars or trucks are then secured before fire-fighters try to get access to the vehicles. During the whole rescue mission each trapped and injured person is attended by a member of the fire-fighters or more commonly by a paramedic. The actions of the fire-fighters and paramedics have to be precisely coordinated because a rough rescue mission can worsen the patients situation.

Modern cars are often fortified to provide more protection. For firefighters this also brings more difficulties to open them with their technical equipment. Therefore, car manufacturers provide special areas that can be cut by firefighters easily without danger. Untriggered airbags have to be localized and prevented from getting triggered.

A train accident is different from a car/van accident. Before approaching the train, the tracks have to be closed to halt further traffic and the overhead lines as well as the train itself has to be grounded. Due to the high amount of passengers many people might need rescue. Rescue dogs may be required to find buried victims.

*C. Mass casualties*

*1) Characteristics and Victims:* Accidents and disasters can cause mass casualties. The injured people are often in dangerous environment and cannot escape and have to be rescued by first responders. The level of injuries of different victims can vary a lot. In fact many people do not have to be injured at all but can be shocked and traumatized so that they also need assistance.

*2) Rescue personnel:* The most important rescue organizations in case of mass casualties are the medical ones. Depending on the size of the accident all available forces a region can provide, are mobilized. Also interregional forces are called in. In cases of a disaster the amount of rescuers involved can be more than 1000.

*3) Strategies:* The main task for fire-fighters is the evacuation of all victims. The latter are brought to a gathering point outside of the danger zone where paramedics and emergency physicians take over them and provide advanced life support. If possible a triage is already made and the patients are brought to the ambulance gathering place where they are transported to the hospitals. If a victim is not stable enough to be transported or there are not enough transport units, it is brought to a treatment point. The purpose of the treatment point is to achieve or maintain transportability of the victim. If the infrastructure is destroyed or nearby hospitals cannot admit more injured people, provisional hospitals have to be established.

The process of triage involves determining the gravity of injury of victims. Based on 4 classifications[5] [19]:

Red    This casualty needs immediate medical attention and has to be transported as soon as possible.
Yellow    This casualty needs medical attention and has to be transported as soon as practically possible.
Green    This casualty requires medical attention after red and yellow classified victims have been treated.
Blue    This casualty has no or small chances of survival. No transport, only observation and if possible administration of analgesics.

Table I: Classification of hazardous materials

| Class | Material | Class | Material |
|---|---|---|---|
| 1 | Explosives | 6 | Toxic and Infectious |
| 2 | Gases (explosive, poisonous) | 7 | Radioactive Substances |
| 3 | Flammable Liquids | 8 | Corrosive Substances |
| 4 | Flammable Solids | 9 | Miscellaneous |
| 5 | Oxidizing | | |

*D. Hazardous materials emergency*

*1) Characteristics:* Most of these emergencies are results of traffic accidents or accidents in industrial buildings. Therefore, all characteristics mentioned for these emergency types are also valid.
Hazardous materials (HazMat) are gases, liquids or solids that can have a damaging impact on people or other living organism. A rough classification is shown in Table I[3].

Some hazardous materials can penetrate or destroy the normal protective clothing of rescuers thus requiring a specialized ones. Moreover, some hazardous materials cannot be avoided from penetrating the human body such as gamma-rays emitted by radioactive materials.

A major risk, especially when gases or infectious materials are involved, is their spread. This can take place due to wind and rain but also through rescue persons or victims that had contact with the hazardous materials and are contaminated. Contamination of the ground or water can have serious impacts on the environment but also directly on the health of people if for example the water system of a city is contaminated.
The Hazardous goods and contaminated area as well as the decontamination materials (water, cloth) have to be collected and disposed. Even small amounts of hazardous material such as radioactive materials can cause problems.

It is important to note, that two different substances that are harmless on their own, can react together and endanger the rescue mission. E.g. some materials react with extinguishing water.

*2) Rescue personnel:* The main work in these scenarios is done by specialized units of fire-fighters. They have proper equipment and protective clothing to handle hazardous materials. A full HazMat team counts up to 30 persons.

*3) Victims:* Victims of HazMat scenarios suffer from the impact of the hazardous material eventually in addition to other injuries caused by accident. Most of the affected victims are contaminated. In some emergency scenarios it is necessary to evacuate the surrounding area.

*4) Strategies [9]:* The first action is rescuing victims. Even if no special equipment is available, units with SCBA enter the contaminated zone. The area with a radius of 50m around the hazardous material is declared as red zone. Entering the red zone is only permitted to specially equipped units. Units operating in the red zone have to rescue victims and avert the risk by proofing, decanting or gathering the hazardous material. Especially for hazardous gas or radioactive materials measuring is very important. If the emergency scenario was caused by an accident, the loaded goods have to be identified. This can happen through a declaration on the vehicle, in the freight documents near the driver or directly through the declaration on the freight itself.
With the knowledge of the hazardous material effective countermeasures can be applied.
At the border of the red zone a decontamination place is established. Units are only allowed to leave the red zone through this place. Rescued victims are also evacuated from the red zone after decontamination.

## VI. SIMILARITIES AND DIFFERENCES

In this section the presented scenarios are examined to find out the similarities and differences that could be useful for the creation of new communication systems.
At first look, the scenarios seem different. A closer look shows however, that all scenarios start with the evacuation of victims. This is the overall main goal of a rescue mission. Anyhow, even if rescuers take risks, they will not bring themselves to unnecessary danger. If a situation seems unfeasible without damage or loss the rescue action is not executed. Self-protection comes first. Therefore rescue units always wear protective clothes and move in groups of minimum of 2 persons. The hierarchy asserts, that approximately eight rescuers have one responsible leader.

The first phase of every emergency is about gathering information. The sooner this happens, the more effective and successful a mission will be. Therefore information and knowledge have to be acquired.
The overall acquired knowledge in an emergency mission is to a great extent versatile. e.g. the remaining air pressure of a self-contained breathing apparatus (SCBA) is not of interest for the rescue mission leader but very important for its carrier.
Therefore, two categories of information can be derived: *Micro information* and *Macro information*.

Micro information is immediately influencing the action of an operating unit or single person. It is not relevant for the overall mission but can anyhow be lifesaving. Micro information can under certain circumstances transform to macro information. e.g. when the air pressure of a fire-fighter's SCBA drops under a critical level.
Macro information is used by leaders to decide over the course of the operation. These decisions are based on an iterative leading process [8] [20] (see Fig. 1).

Basic information needed for a decision process can be found in Table II [8].

---

[3]http://www.tes.bam.de/de/regelwerke/klassifizierung/index.htm

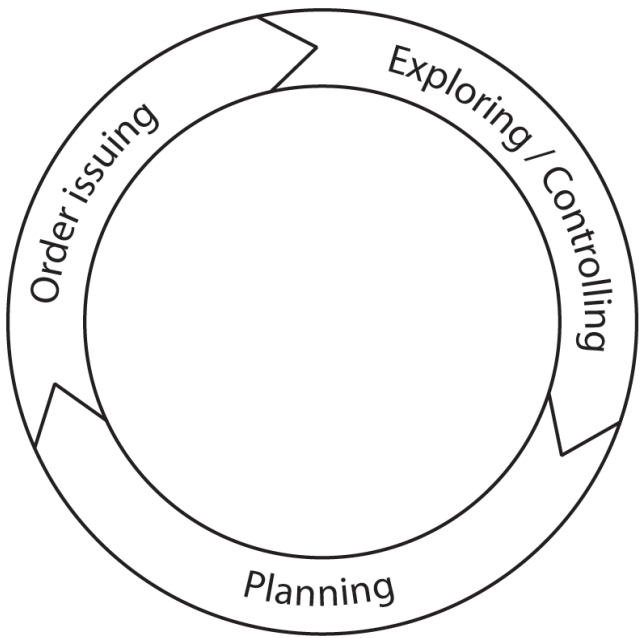

Figure 1: The iterative leading process.

Table II: Information for decision process

| Damage | Damaged object | Damage scope |
|---|---|---|
| Type | Type | Humans |
| Cause | Size | Animals |
|  | Material | Environment |
|  | Construction |  |
|  | Surrounding |  |

During the whole emergency mission position information of rescue units and victims are important. For an indoor rescue mission, the position of missing people is of utmost importance.[2].

## VII. OPEN TOPICS

From the scenarios in section V and the interviews a range of different improvement possibilities came into light. First of all, a basic communication channel for all participants of a rescue mission is required. There has already been a lot of research in the field of mobile middleware using ad-hoc networks. e.g. the MIDAS project[21] and its predecessor [6] or [22]. However, no system has been successfully tested in a large scale emergency scenario. The acceptance of new communication system from the user perspective was also not evaluated. Is the knowledge about efficient usability transferable to an emergency device, and can it be handled under time pressure, stress and in a harmful environment?
If a stable communication system exists, this could be used by many future technologies supporting the emergency team collecting information. In cases of conflagration a big danger for fire-fighters is to get lost and trapped in a smoke filled buildings and then running out of breathable air.
Therefore, indoor locating of the rescuers is an important problem that has not been solved yet. Different approaches such as tracing the rescue units with wearable sensors[4] or building sensor lifelines by automatically dropping sensor nodes[23].
Approaches involving participatory sensing, such as the inclusion of mobile devices like smartphones of victims can help in the collection of information through their sensors. By including the victims mobile devices, new possibilities of locating them arise. With the progresses made in context aware homes, the context awareness of buildings could also be used by the firefighters to locate inhabitants. The context awareness could not only provide information about the location but also about the state of the emergency site. Temperature sensors, motion sensors or video cameras accessible by the firefighters could provide a wide range of useful life information. Not only this, but passive information could also be provided by a context aware environment such as floor plans, points of danger like as stored chemicals, power transformer of solar collectors or dangerous machines.

To prevent accidents of rescue units through exhaustion, heat, gas and other harmful aspects intelligent clothing could be used. These sensor equipped clothes monitor vital signs such as the heart rate and body temperature but also surrounding parameters such as gas composition[4] or the air pressure of the SCBA. All sensors are part of a personal area network. The energy supply can be realized though a classical battery system but also trough energy producing parts in the clothing itself.

To gather information about inhabitants that could potentially be missed, an autonomously started deep Internet research taking also into account social networks could provide important information about aspects such as age, handicap or other critical information for a rescue team.

In cases of traffic accidents the principles of context awareness can also be applied. Context aware vehicles would provide the rescue units important information. This information could be specific rescue instructions (dismantling points, position of airbags), state information like the fuel temperature or identification of the passengers. The loaded goods can also contain hazardous materials. Providing the exact information about load to the emergency helpers before they start operating would also be a benefit because more precise actions could be induced earlier.
If the traffic accident affects a larger amount of people, participatory sensing can also provide supplementary information such as the position of casualties.

Mass casualties require a triage. Different research projects have investigated in improving the triage system. Most of them require a stable Internet connection because their data is stored on a central server. In cases of mass casualties through disasters it is not unlikely, that the complete public communication infrastructure is damaged and not working. Decentralizing the triage process on a peer to peer level would at least enable the local use of these systems on the emergency scene.

[4]http://www.wearitatwork.com

To get an overview of lager emergency scenarios, aerial support in form of unmanned aerial vehicles taking pictures of the scene can be deployed. In cases of disrupted communication infrastructure, communication devices such as routers could be positioned high above the incident (e.g. ground attached balloon) and build a cell.

For larger emergencies involving several rescue teams a command stand has to be established. Innovative software could support the leading process. If all rescue members can be located and a stable communication infrastructure is established all information from the emergency area can be streamed into this command stand in real time. An easy to use interface should provide all relative information in an easy and accessible way without distracting the operator. The software could benefit from the development in computer games. An interface like a strategy game would provide an overview of all units and their movement. By selecting different objects/persons on the screen additional information could be displayed. Complex orders like the division of the emergency area into zones and the allocation of rescue teams to these zones or repositioning of vehicles could be created by click actions in the software and then be send to the responsible section leaders and drivers.

The overall availability of information could also be used by the rescuers' personal mobile devices to warn them before entering forbidden areas like the red zone in a HazMat scenario.

## VIII. Conclusion

In this paper a classification of emergency scenarios was presented. Emergency scenarios are composed of one or more hazardous events. The unpredictability requires fast and creative reaction from the first responders. It could be shown that independent of the occurred events the pattern of emergency response is always the same: Information collection » Planning » Order issuing/Acting. The primary task is rescuing victims.

Based on extensive interviews with professional rescuers and fire-fighters, requirements for new communication systems were listed and described. It can be said, that there is a high potential for developing new innovative systems to support first responders. But the requirements are strict and new systems have to be stable and user friendly to get accepted by the conservative rescuers.
New software and hardware will primarily collect data or provide the communication. Wireless Ad Hoc networks seem to be the right choice because they are self-configuring and need no infrastructure.
Sensor networks provide the largest field for innovation. They will become ubiquitous on an emergency site.

## References

[1] N. C. Sanderson, "Network wide information sharing in rescue and emergency situations," Ph.D. dissertation, University of Oslo, 2008. [Online]. Available: http://folk.uio.no/noruns/thesisFinal.pdf
[2] A. Ruhs, December 2010. [Online]. Available: http://www.feuerwehr-frankfurt.de/
[3] J. Rönnfeldt, 01 2011. [Online]. Available: http://www.darmstadt.de/leben-in-darmstadt/sicherheit/index.htm
[4] Schäfer, 01 2011. [Online]. Available: http://merck.de/
[5] M. D. Gennaro, 01 2011. [Online]. Available: http://www.drkfrankfurt.de
[6] N. Sanderson, K. S. Skjelsvik, O. V. Drugan, M. Puzar, V. Goebel, E. Munthe-Kaas, and T. Plagemann, "Developing mobile middleware - an analysis of rescue and emergency operations," Department of informatics, University of Oslo, Norway, Tech. Rep. Research Report no.358, jun 2007.
[7] *World Disasters Report 2005*. International Federation of Red Cross and Red Crescent Societies (IFRC), 2005. [Online]. Available: http://www.ifrc.org/publicat/wdr2005
[8] *Feuerwehr-Dienstvorschrift 100*, book, Ausschuss Feuerwehrangelegenheiten, Katastrophenschutz und zivile Verteidigung (AFKzV) Std., March 1999. [Online]. Available: http://www.hlfs.hessen.de/irj/HLFS_Internet?cid=41bc283980d84ca9d21430e3257af00b
[9] *Feuerwehr-Dienstvorschrift 500 - Einheiten im ABC-Einsatz*, book, Ausschuss Feuerwehrangelegenheiten, Katastrophenschutz und zivile Verteidigung (AFKzV) Std., August 2004. [Online]. Available: http://www.hlfs.hessen.de/irj/HLFS_Internet?cid=41bc283980d84ca9d21430e3257af00b
[10] D. Feuerwehrverband, "Feuerwehr in zahlen," Tech. Rep., December 2007. [Online]. Available: http://www.dfv.org/fileadmin/dfv/Dateien/Presse/Praesentation__Feuerwehr_in_Zahlen_.pdf
[11] D. R. K. (DRK). [Online]. Available: http://www.drk.de/
[12] Malteser. [Online]. Available: http://www.malteser.de/
[13] Johanniter-Unfall-Hilfe. [Online]. Available: http://www.johanniter.de/die-johanniter/johanniter-unfall-hilfe/
[14] C. Linde, *Aufbau und Technik des digitalen BOS-Funks*. Franzis Verlag GmbH, 2008.
[15] *Katastrophenschutz in Hessen*, Hessisches Ministerium des Innern und für Sport Std., 2011. [Online]. Available: http://www.hmdi.hessen.de/irj/servlet/prt/portal/prtroot/slimp.CMReader/HMdI_15/HMdI_Internet/med/611/6115e6cf-f30f-c21f-012f-31e2389e4818,22222222-2222-2222-2222-222222222222,true
[16] *Unfallverhütungsvorschrift Feuerwehren (GUV-V C53)*, Gesetzliche Unfallversicherung Std. [Online]. Available: http://publikationen.dguv.de/dguv/pdf/10002/V_C53.pdf
[17] *Einsatzstichworte*, Staatsanzeiger für das Land Hessen, Hessisches Ministerium des Innern und für Sport Std., 08 2009.
[18] *Qualitätskriterien für die Bedarfsplanung von Feuerwehren in Städten*, Arbeitsgemeinschaft der Leiter der Berufsfeuerwehren in der Bundesrepublik Deutschland Std. [Online]. Available: http://agbf.de/joomla/images/stories/bf/aka/qualitaetskriterien_fuer_bedarfsplanung_von_feuerwehren_in_staedten.pdf.pdf
[19] *Konzept zur überörtlichen Hilfe bei MANV*, Arbeitsgruppe der Hilfsorganisationen im Bundesamt für Bevölkerungsschutz und Katastrophenhilfe Std. [Online]. Available: http://www.bbk.bund.de/cln_027/nn_402322/SharedDocs/Publikationen/Publikation_20KatMed/Hilfekonzept__bei__MANV,templateId=raw,property=publicationFile.pdf/Hilfekonzept_bei_MANV.pdf
[20] S. Mehrotra, "Project rescue: challenges in responding to the unexpected," in *Proceedings of SPIE*, San Jose, CA, USA, 2003, pp. 179–192. [Online]. Available: http://spie.org/x648.html?product_id=537805
[21] E. Munthe-Kaas, A. Johannessen, M. Puzar, and T. Plagemann, "Information sharing in mobile ad-hoc networks: Metadata management in the midas dataspace," in *Tenth International Conference on Mobile Data Management: Systems, Services and Middleware*, 2009.
[22] R. B. Dilmaghani and R. R. Rao, "Hybrid wireless mesh network with application to emergency scenarios," *JOURNAL OF SOFTWARE2008*, vol. 3, No. 2, pp. 52–60, February.

[23] M. Klann, T. Riedel, H. Gellersen, C. Fischer, G. Pirkl, K. Kunze, M. Beuster, M. Beigl, O. Visser, and M. Gerling, "Lifenet: an ad-hoc sensor network and wearable system to provide firefighters with navigation support." [Online]. Available: http://citeseerx.ist.psu.edu/viewdoc/download?doi=10.1.1.103.5633&rep=rep1&type=pdf

Table III: Classification of emergency scenarios

| Emergency Scenario | Characteristics | Victims | Strategy |
|---|---|---|---|
| Conflagration / Fire | <ul><li>Predominant in urban areas / buildings</li><li>Danger of explosion (Gas pipes / tanks)</li><li>Danger of collapsing building</li><li>Danger of electricity</li><li>Poisonous smoke</li><li>Loss of orientation</li><li>Enormous heat</li><li>Flash over</li><li>Spreads easily</li></ul> | <ul><li>Usually leave the dangerous area</li><li>Can get lost in unfamiliar environment</li><li>At night smoke surprises victims → No escape</li><li>Children tend to hide from smoke</li><li>Once evacuated, not likely that missing people evacuate by their own</li></ul> | <ul><li>Exploring</li><li>All available rescue forces sent inside the building</li><li>Rescue units move on the ground</li><li>Use their hands to find victims</li><li>Infrared cameras used</li><li>Coordinate their positions over radio</li><li>Leader has to remember position and way</li><li>If more forces arrive → area separated into zones</li><li>Command stand is established - multiple communication channels</li></ul> |
| Traffic accident | <ul><li>Highest danger through on-going traffic</li><li>Danger of ignition and explosion</li><li>Dangerous load: Chemicals, gas or fuel</li><li>Accident in urban areas → evacuation</li><li>Danger through Airbags</li><li>Train: Danger through Electricity</li><li>Mountainous area, tunnels</li></ul> | <ul><li>Usually injured and under shock.</li><li>Loss of blood</li><li>Trapped in their vehicles.</li><li>More time available to work in a more controlled and delicate way.</li><li>High amount of victims. Different levels of injury.</li></ul> | <ul><li>First action: blocking traffic</li><li>Trapped and injured victims are attended by paramedics</li><li>Vehicles are secured and then opened</li><li>Untriggered airbags are secured</li><li>Constant coordination between paramedics and fire-fighters</li><li>Before approaching → Grounding overhead lines and train</li><li>Disaster → Mass casualties</li></ul> |
| Mass casualties | <ul><li>Injured people in dangerous environment</li><li>Psychological stress</li><li>Panic</li><li>Trauma</li></ul> | | <ul><li>Fire-fighters evacuate victims to gathering point outside of the danger zone</li><li>Paramedics take them over → advanced life support</li><li>Triage</li><li>If stable → ambulance gathering point → transport to hospital</li><li>If unstable → treatment area</li></ul> |

| Hazardous materials | - gases, liquids or solids with damaging impact
- penetrate or destroy the normal protective clothing
- Spreading through wind, rain, rescue personnel or victims
- Impact on environment
- Some HazMat react with water | - suffer under impact of hazardous material
- contaminated
- maybe necessary to evacuate surrounding area | - Rescuing victims
- Radius of 50m around the hazardous material is declared as red zone.
- Only permitted to specially equipped units
- Proofing, decanting or gathering the hazardous material
- Identification of freight
- At the border of the red zone → decontamination place
- Only allowed to leave the red zone through this place
- Victims get also decontaminated |
|---|---|---|---|
| Forest fire | - Large scale
- No water available to extinguish fire.
- Fast spreading
- Embers → reinflammation | | - Establishing water supply → commuting fire trucks or long distance hose lines
- Aerial support (water bomber)
- Creating fire break lines |
| Flood | - Disaster
- Large scale
- Infrastructure destroyed
- Danger of epidemic | - Many people affected
- People loose homes
- Evacuation of large region | - National/International help required
- Providing shelter, medical treatment, nutrition and drinkable water for rescuers and population
- Building levees
- Evacuating population with boats or helicopters
- Draining flooded buildings and places
- Rebuilding infrastructure
- Long operating time
- High coordination effort |
| Earthquake | - Disaster
- Large scale
- Infrastructure destroyed
- Danger of epidemic
- Danger of further collapsing buildings | - Many people affected
- People loose homes
- People trapped and buried in collapsed buildings
- Many people injured and dead | - National/International help required
- Providing shelter, medical treatment, nutrition and drinkable water for rescuers and population
- Search for victims → rescue dogs
- Rebuilding infrastructure
- Long operating time
- High coordination effort |